\documentstyle[11lomcon,cite,psfig]{article}
\begin{document}

\title{GRAVITATIONAL MICROLENSING AND DARK MATTER IN OUR GALAXY:
10 YEARS LATER}

\author{A.F. ZAKHAROV\footnote{e-mail: zakharov@itep.ru}
}

\address{Institute of Theoretical and Experimental Physics,
 117259,  B. Cheremushkinskaya, 25, Moscow, Russia
}
\address{Astro Space Centre of Lebedev Physics Institute, Moscow, Russia}


\maketitle\abstracts{ Foundations of standard theory of
microlensing are described, namely we consider microlensing  stars
in Galactic bulge, the Magellanic Clouds or other nearby galaxies.
We suppose that gravitational microlenses lie between an Earth
observer and these stars. Criteria of an identification of
microlensing events are discussed. We also consider such
microlensing events which do not satisfy these criteria
(non-symmetrical light curves, chromatic effects, polarization
effects). We describe  results of MACHO collaboration observations
towards the  Large Magellanic Cloud (LMC) and the Galactic bulge.
Results of EROS observations towards the LMC and OGLE observations
towards the Galactic bulge are also presented. Future microlensing
searches  are discussed.}

A standard microlens model is based on a simple approximation of a
point mass for a gravitational microlens. Gravitational lensing
(gravitational focusing) results from the effect of light bending
by a gravitating body (the phenomenon was discussed by I.~Newton,
but in the framework of Newtonian gravity a formal derivation of
the light bending angle was published by J.~Soldner
\cite{Soldner04}).

In the framework of general relativity (GR) using
 a weak gravitational field approximation the correct bending angle
is described by the following expression derived by Einstein in
1915 just after his formulation of GR
\begin{eqnarray} \delta
\varphi=-\frac{4GM_*}{c^2 p}.  \label{eqs5}
\end{eqnarray}
The derivation of the famous Einstein's formulae for the bending
angle of light rays in gravitational field of a point mass $M_*$
is practically in all monographs and textbooks on general
relativity and gravity theory (see, for example books
\cite{Landau75,Moller72}).

The law was firstly confirmed by Sir A. Eddington for observations
of light ray bend by the Solar gravitational field near its
surface. The angle is equal to $1.75''$, therefore Einstein
prediction was confirmed by observations very soon after its
discovery.

The gravitational lens effect is a formation of several images
instead of one (see details in \cite{Zakharov97,Zakharov98_2}). We
have two images for a point lens model (Schwarzschild lens model).
The total square  of the two images is larger than a source
square. The ratio of these two squares is called gravitational
lens amplification $A$. That is a reason to call gravitational
lensing as gravitational focusing. The angular distance between
two images is about angular size of so-called Einstein's cone. The
angular size of Einstein's cone is proportional to the lens mass
divided by the distance between a lens and an observer. Therefore,
if we consider a gravitational lens with typical galactic mass and
a typical galactic distance between a gravitational lens and an
observer then the angular distance between images will be about
few angular seconds; if we suppose that a gravitational lens has a
solar mass and a distance between the lens and an observer is
about several kiloparsecs then an angular distance between images
will be about angular millisecond.

If a separation angle is $\sim 1''$, then one may observe two
images in optical band although this problem is a complex  one,
but one cannot observe directly two images by Earth's observer in
the optical band if a separation angle is $\sim 0.001''$.
Therefore, the microlensing effect is observed on changing of a
luminosity of a source $S$.\footnote{Since the angle is very small
Einstein and Chwolson thought that gravitational lens effect could
not be detectable if sources and lenses are stars. Now there are
chances to measure such angles in IR band therefore there is a
giant development of observational facilities.}

If the source $S$ lies on the boundary of the Einstein cone, then
we have  $A=1.34$. Note, that the total time of crossing  the
Einstein cone is $T_0$. Sometimes the microlensing time is defined
as a half of $T_0$ we suppose that $D_{d}<D_{ds}$ (here we assume
that $D_{ds}$ is the distance from the source $S$ to the lens $D$;
$D_{d}$ is the distance from the lens $D$ to the observer $O$;
$D_{s}$ is the distance from the source $S$ to the observer $O$)
$$
T_0=3.5~ months \cdot \sqrt{\frac{M}{M_{\odot}} \frac{D_{d}}{10\, kpc}}
\cdot \frac{300\, km/s}{v},
$$
where $v$ is the perpendicular component of a velocity of a dark
body. If we suppose that the perpendicular component of a velocity
of a dark body is equal to $\sim 300$ km/s (that is a typical
stellar velocity in Galaxy), then a typical time of crossing
Einstein cone is about 3.5 months. Thus, a luminosity of a source
$S$ is changed with the time.

We will give numerical estimations for parameters of the
microlensing effect. If the distance between a dark body and the
Sun is equal to $\sim 10$~kpc, then the angular size of Einstein
cone of the dark body with a solar mass is equal to  $\sim
0.001''$ or the linear size of Einstein cone is equal to about 10
astronomical units. It is clear that since  typical distances
between two images are about Einstein diameters therefore is very
difficult  to resolve the images by ground based telescopes at
least in an optical band. It was a reason that both Einstein and
Chwolson thought if gravitational lenses and sources are stars
then separation angle is very small to be detectable. However,
recently, a direct method to measure Einstein angle $\phi_E$ was
proposed to resolve double images generated by microlensing with
an optical interferometer (say VLTI) \cite{Delpl01}(see also
\cite{Pacz03} for a discussion). Moreover, it is plan to launch
astrometrical space probe, American
SIM\footnote{http://sim.jpl.nasa.gov/whatis/} and European
GAIA\footnote{http://astro.estec.esa.nl/GAIA}, these instruments
will have precisions about 10 micro arc seconds and could
determine Einstein radii for any microlensing events.

Astrometric microlensing or motions of visible images due to
influence of a gravitational field of  microlenses was analyzed in
number of papers
 \cite{Hog95,Walker95,Miyamoto95,Sazhin96,Sazhin98,Pacz98,Boden98,Honma01,Honma02,Takahashi03,Lewis98,Treyer03},
although light bending in gravitational field was discussed by
I.~Newton (actually that is the same effect but authors presented
detailed analysis and pushed new ideas to use the phenomenon to
detect even invisible astronomical objects by shifts of images for
background sources). An optical depth of microlensing for
distant quasars was discussed for different locations of
microlenses  (see, for example, \cite{Zakharov04} and references therein.

For observations of extragalactic gravitational lens a typical
time for changes of light curve is very long
 ($\sim 10^{5}\!$ years) for its direct observations.
Therefore, extragalactic gravitational lenses are discovered and
observed by resolving different optical components (images) since
typical angular distances between images are about some angular
seconds because of a great mass of a gravitational lens. If a
gravitational lens is a galaxy cluster then the angular distances
between images may be about several minutes. For an identification
of gravitational lenses, observers compare typical features and
spectra of different images. It is clear that one cannot to
resolve different components during microlensing but it is
possible to get and analyze a light curve in different spectral
bands.

One of the basic criterion for microlensing event identification
is the symmetry of a light curve. If we consider a spherically
symmetric gravitational field of a lens, a point source and a
short duration of microlensing event then the statement about the
symmetry of a light curve will be a strong mathematical
conclusion, but if we consider a more complicated distribution of
a gravitational field lens or an extensive light source then some
deviations of symmetric light curves may be observed and (or) the
microlensing effect may be chromatic
\cite{Zakharov97,Zakharov98_2}.

More than 70 years ago it was found that densities of visible
matter is about 10\% of total density in galactic halos (the
invisible is called as dark matter (DM)
\cite{Oort32,Zwicky33})\footnote{Now it is known that the matter
density (in critical density units) is $\Omega_m=0.3$ (including
baryonic matter $\Omega_b \approx 0.05-0.04$, but luminous matter
$\Omega_{\rm lum} \approx 0.001$), $\Lambda$-term density
$\Omega_\Lambda=0.7$.} Thus baryonic density is a small fraction
of total density of the Universe. Probably galactic halos is
"natural" places to store not only baryonic DM, but non-baryonic
DM also. If DM forms objects with masses in the range
$[10^{-5},10]M_\odot$ microlensing could help to detect such
objects. Thus, before intensive microlensing searches it was a
dream that microlensing investigations could help us to solve DM
problem for Galactic halo at least.

For the first time a possibility to discover microlensing using
observations of star light curves was discussed in the paper by
Byalko in 1969 \cite{Byalko69}. Systematic searches of dark matter
using typical variations of light curves of individual stars from
millions observable stars started after Paczynski's discussion  of
the halo dark matter discovery using monitoring stars from Large
Magellanic Cloud (LMC) \cite{Paczynski86}. We remark that in the
beginning of the nineties new computer and technical possibilities
providing the storage and processing of huge volume of
observational data were appeared and it promoted at the rapid
realization of Paczynski's proposal. Griest  suggested to call the
microlenses as Machos (Massive Astrophysical Compact Halo Objects)
\cite{Griest91}. Besides, MACHO is the name of the project of
observations of the US-English-Australian collaboration which
observed the LMC and Galactic bulge using 1.3 m telescope of Mount
Stromlo observatory in Australia.\footnote{MACHO stopped since end
1999.}

The first papers about the microlensing discovery were published
by the MACHO collaboration \cite{Alcock93} and the French
collaboration EROS (Exp\'erience de Recherche d'Objets Sombres)
\cite{Aubourg93}.\footnote {EROS experiment stopped in 2002
\cite{Moniez01}.}

First papers about the microlensing discovery toward Galactic
bulge were published by US-Polish collaboration (Optical
Gravitational Lens Experiment), which used 1.3 m telescope at Las
Campanas Observatory. Since June 2001, after second major hardware
upgrade OGLE entered into its third phase, OGLE III as a result
the collaboration observes more than 200 millions stars observed
regularly once every 1 -- 3 nights. Last two years OGLE III
detected more than four hundreds microlensing event candidates
each year
\cite{Udalski2002}.\footnote{http://www.astrouw.edu.pl/~ogle/ogle3/ews/ews/html}

MOA (Microlensing Observations in Astrophysics) is collaboration
involving astronomers from Japan and New Zealand
\cite{Bond01,Skuljan03}.\footnote{http://www/roe.ac.uk/\%7Eiab/alert/alert/alert/html}

To investigate Macho distribution in another direction one could
use searches toward M31 (Andromeda) Galaxy lying at 725 kpc (it is
the closest galaxy for an observer in the Northern hemisphere). In
 nineties two collaborations AGAPE (Andromeda Gravitational
 Amplification Pixel Experiment, Pic du Midi, France)\footnote{New collaboration, POINT-AGAPE started in 1999 and uses
INT (2.5 v Isaac Newton Telescope)\cite{Kerins01}.}
 and VATT
 started to monitor pixels instead of individual stars
 \cite{Moniez01,LeDu01}. These teams reported about discoveries of
 several microlensing event candidates.

The event corresponding to microlensing may be characterized
by the following main features, which allow to distinguish
the microlensing event and a stellar variability
\cite{Roulet97,Zakharov97,Zakh02a}.  \\
\begin{itemize}

\item Since the microlensing events have a very small probability,
the events should never repeat for the same star. The stellar
variability is connected usually with periodic (or quasi-periodic)
events of the fixed star. \item In the framework of a simple model
of microlensing when a point source is considered, the
microlensing effect must be achromatic (deviations from
achromaticity for non-point source were considered, for example in
the paper by Bogdanov \& Cherepashchuk \cite{Bogdanov95}), but the
proper change of luminosity star is connected usually with the
temperature changes and thus the light curve depends on a colour.
\item The light curves of microlensing events are symmetric, but
the light curves of variable stars are usually asymmetric (often
they demonstrate the rapid growth before the peak and the slow
decrease after the peak of a luminosity).
\item Observations of
microlensing events are interpreted quite well by the simple
theoretical model, but some microlensing events are interpreted by
more complicated model in which one can take into account that a
source (or a microlens) is a binary system, a source has
non-vanishing size, the parallax effect may take place.
\end{itemize}

\begin{figure}[!t]
\vspace{9.2cm} \includegraphics{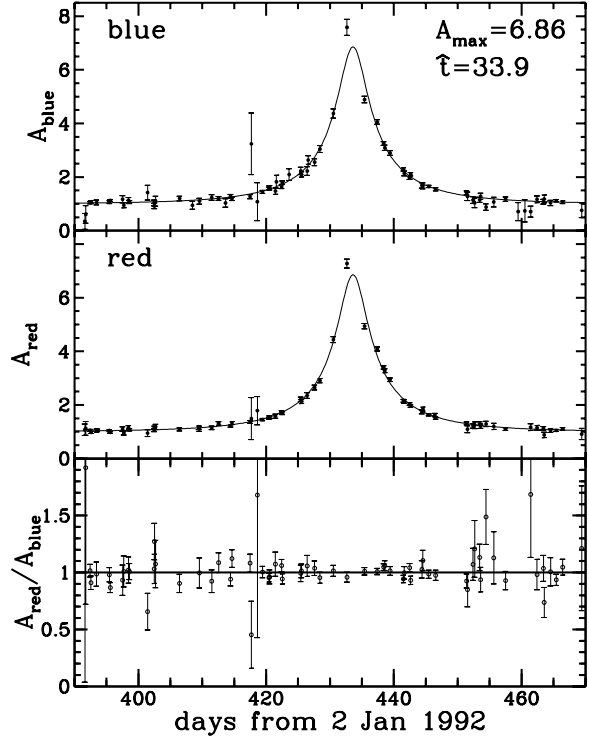} \caption{The first microlensing event
which was detected by the MACHO collaboration during microlensing
searches towards LMC \cite{Alcock93}.}
\label{gold}
\end{figure}

The typical features of the light curve of the first microlensing
event observed by the MACHO collaboration in the LMC  are shown in
Fig. \ref{gold}, where the light curves are shown for two spectral
bands (a more recent MACHO fit to the observed amplification of
this event gives $A_{\rm max}=7.2$). The light curve (in two
bands) is fitted by the simple model well enough, but the ratio of
luminosities for the bands is shown in the lower panel of figure
(the ratio shape is adjusted with the event achromaticity).
However, one can note that near the maximal observable luminosity
the theoretical curve  fits the data of observations not very
well.

Now one can carry out accurate testing  the achromaticity
and moreover the stability of the source spectrum
during a microlensing event with the Early Warning systems
implemented both by the MACHO  and OGLE
collaborations. This allows  one to study the source properties
using large telescopes and to organize
intense follow-up studies of light curves
using telescope network around the globe.

In addition to the typical properties of individual microlensing
events, Roulet and Mollerach note that the population of observed
events should have the following statistical properties
\cite{Roulet97,Zakharov97}.

\begin{itemize}

\item Unlike a star variability microlensing events should happen
with the same probability for any kind of star therefore the
distribution of microlensing events should correspond to the
distribution of observed stars in the color-magnitude
diagrams.\footnote{However, Roulet and Mollerach  noted that for
observations in the bulge since observed stars have non-negligible
spread along the line of sight, the optical depth is significantly
larger for the star lying behind the bulge, thus the lensing
probabilities should increase for the fainter
stars\cite{Roulet97}.}

\item The distribution of the maximal amplification factor $A_{\rm
max}$ should correspond to a uniform distribution of the variable
$u_{\rm min}=1/b$ ($b$ is the dimensionless impact parameter).

\item The distributions of the amplification $A_{\rm max}$ and the
microlensing event time $T$ should be uncorrelated.

\end{itemize}

Since for the microlens searches one can monitor several million
stars for several years, the ongoing searches have focused on two
targets: a) stars in the Large and Small Magellanic Clouds (LMC
and SMC) which are the nearest galaxies having lines of sight
which go out of the Galactic plane and well across the halo; b)
stars in the Galactic bulge which allow to test the distribution
of lenses near to the Galactic plane.\footnote{In this paper we do
not discuss microlensing for distant quasars.}

Let us cite well established results of microlensing searches and
discuss the questions for which we have now different answers
which do not contradict to the observational data. Now it is
generally recognized that the microlensing searches towards the
Galactic bulge or nearby galaxies are very important for solutions
of a lot of problems in astronomy and cosmology. As Paczynski
noted, the most important is the consensus that the microlensing
phenomenon has been discovered \cite{Paczynski96}. Now it is
impossible to tell which part of the microlensing event candidates
is actually connected with the effect since probably there are
some variable stars among the event candidates, it could be
stellar variability of an unknown kind.\footnote{ The microlensing
event candidates proposed early by the EROS collaboration ( \#1
and \#2) and by the MACHO collaboration (\#2 and \#3) are
considered now as the evidence of a stellar variability
\cite{Paczynski96}.}

\begin{enumerate}
\item Observed light curves are  achromatic and their shapes are
interpreted by simple theoretical expressions very well, however,
there is not complete consent about "very well interpretation"
since even for the event candidate MACHO \# 1 the authors of the
discovery proposed two fits. Dominik and Hirshfeld  suggested that
the event could be fitted perfectly in the framework of the
binary lens model \cite{Dominik94,Dominik95}, but Gurevich et al.
assumed that the microlensing event candidate could be caused by a
non-compact microlens \cite{Gurevich96}.\footnote{Microlensing by
non-compact objects considered also in papers
\cite{Zakharov98_1,Zakharov99,Zakharov01a,Zakharov01b,Zakharov96a,Zakharov96b}.}

\item
As expected, binary lenses have been detected
and the observed rate of the events correspond to expected value.

\item
As expected, the parallax effect has been detected.

\item
Since the observed optical depth
is essentially greater than the estimated value,
the independent confirmation of the Galactic bar existence was done.

\item Using photometric observations of the caustic-crossing
binary lens microlensing event EROS BLG-2000-5, PLANET
collaboration reported about the first microlens mass
determination, namely the masses of these components are 0.35
$M_\odot$ and 0.262 $M_\odot$ and the lens lies within 2.6 kpc of
the Sun \cite{An02}.

\item Bennett et al.  discovered gravitational microlensing events
due to stellar mass black holes \cite{Bennett02}. The lenses for
events MACHO-96-BLG-5 and MACHO-96-BLG-6 are the most massive,
with mass estimates $M/M_\odot=6^{+10}_{-3}$ and
$M/M_\odot=6^{+7}_{-3}$, respectively.

\end{enumerate}

Now the following results are generally accepted:

\begin{enumerate}
\item The optical depth towards the Galactic bulge is equal to $
\sim 3 \times 10^{-6} $, so it is larger than the estimated value
\cite{Alcock00a}.

\item

Analysis of 5.7 years of photometry on 11.9 million stars in LMC
by MACHO collaboration reveals 13 -- 17 microlensing events
\cite{Alcock00b} (recent results of the MACHO collaboration on
could find in \cite{Popow03}). The optical depth towards the LMC
is equal to $\tau(2 < \hat{t} < 400 {\rm~ days})
=1.2^{+0.4}_{-0.3} \times 10^{-7} $, so, it is smaller than the
estimated  value. The maximum likelihood analysis gives a MACHO
halo fraction f=0.2. Alcock et al. (2000b) gives also estimates of
the following probabilities $P (0.08 < f <0.5)=0.95$ and $P(f=1) <
0.05.$ The most likely MACHO mass $M \in [0.15, 0.9] M_\odot$,
depending on the halo model and total mass in MACHOs out 50 kpc is
found to be $9^{+4}_{-3} \times 10^{10} M_\odot$ EROS
collaboration gives a consistent conclusion, namely, this group
estimates the following probability $P (M \in [10^{-7},
1]M_\odot~\&~f>0.4) <~ 0.05$ \cite{Lasserre00,Lasserre01}.
However, these conclusions are based on assumptions about mass and
spacial distributions of microlenses but generally speaking these
distributions are still unknown.

\end{enumerate}

However there are different suggestions (which are not
contradicted to the observational data) about the following issues
\cite{Paczynski96}:

{\it What is the location of objects which dominate microlensing
observed towards the Galactic bulge?}

{\it Where are the most microlenses for searches towards LMC?} The
microlenses may be in the Galactic disk, Galactic halo, the LMC
halo or in the LMC itself. {\it Are the microlenses stellar mass
objects or are they substellar brown dwarfs?}

{\it What fraction of microlensing events is caused by
 binary lenses?}

{\it What fraction of microlensing events is connected
with binary sources?}

Paczynski suggested that we shall have definite answers for some
presented issues after some years and since the optical depth
towards the Galactic bulge is essentially greater than the optical
depth towards the LMC, we shall have more information about the
lens distribution towards the Galactic bulge, however, probably,
some problems in theoretical interpretation will appear after
detections of new microlensing event candidates
\cite{Paczynski96}.

The main result of the microlensing searches is that
the effect predicted theoretically has been confirmed.
This is one of the most important astronomical discoveries.

When new observational data would be collected and the processing
methods would be perfected, probably some microlensing event
candidates lost their status, but perhaps new microlensing event
candidates would be extracted among analyzed observational data.
So, the general conclusion may be done. The very important
astronomical phenomenon was discovered, but some quantitative
parameters of microlensing will be specified in future. However,
the problem about 80\% of DM in the halo of our Galaxy is still
open (10 years ago people believe that microlensing could give an
answer for this problem). Thus, describing the present status
Kerins wrote adequately that now we have "Machos and clouds of
uncertainty" \cite{Kerins01}.

I thank prof. A.I.~Studenikin for his kind invitation to present
this contribution at the  XI Lomonosov Conference on Elementary
Particle Physics.


\end{document}